\documentstyle[preprint,aps,eqsecnum]{revtex}
\tightenlines
\def\m@thcombine#1#2{%
  \setbox0=\hbox{$#1$}
  \setbox1=\hbox{$#2$}
  \ifdim\wd0>\wd1
    \setbox0=\hbox to\wd1{\hss\box0\hss}
  \else
    \setbox1=\hbox to\wd0{\hss\box1\hss}
  \fi
  \mathop{\vcenter{
    \offinterlineskip\box0\box1}}}
\def\lesim{\m@thcombine<\sim}
\def\gesim{\m@thcombine>\sim}

\begin{document}

\draft
\title{STRUCTURE OF THE YANG-MILLS VACUUM IN THE \\ ZERO MODES ENHANCEMENT QUANTUM MODEL }

\author{V. Gogohia }

\address{HAS, CRIP, RMKI, Theory Division, P.O.B. 49, H-1525, \\
          Budapest 114, Hungary. Email address: gogohia@rmki.kfki.hu }

\maketitle

\begin{abstract}
We have formulated new quantum model of the QCD vacuum using the effective potential approach for composite operators. It is based on the existence and importance
of such kind of the nonperturbative, topologically nontrivial excitations of gluon field configurations, which can be effectively correctly described by the $q^{-4}$-type behaviour of the full gluon propagator in the deep infrared domain. The ultraviolet part of the full gluon propagator was approximated by the asymptotic freedom to-leading order perturbative logarithm term of the running coupling constant. Despite the vacuum energy density remains badly divergent, we have formulated a method how to establish a finite (in the ultraviolet limit) relation between the two scale parameters of our model. We have expressed the asymptotic scale parameter as $pure \ number$ times the nonperturbative scale, which is inevitably contained in any realistic Ansatz for the
full gluon propagator.
\end{abstract}


\vfill

\eject

\section{Introduction}

The modern theory of strong interactions, quantum chromodynamics
(QCD) [1] is intrinsically ultraviolet (UV) divergent theory
because of asymptotic freedom [2,3] which is due to
self-interaction of massless gluons only (non-Abelian character
of QCD).  It implies that any exact solution to QCD should become
asymptotically free in the deep UV limit. This means that the
perturbative UV divergences will be inevitably encountered in
order to calculate any physical observable in this theory. Thus
the existence of the UV divergences in QCD does not depend on the
method of solution because behind these divergences is a physical
phenomenon - asymptotic freedom. In other words, there is no way
to escape from the UV divergences in QCD by implementing some
sophisticated (other than perturbation theory) method of
solution. Since divergences are physical (not artificial due to a
method of solution), a mass scale parameter appears as a result
of the UV renormalization program in order to render the theory
finite in the UV limit. In QCD quark masses are not observable,
so the UV divergences are to be absorbed into the scale parameter
which precisely becomes finite in the UV limit. Instead of the
dimensionless fixed coupling constant (like in QED, see a brief
discussion below) we are left with the above-mentioned
dimensional asymptotic scale parameter, $\Lambda_{QCD}$ and
therefore dimensionless coupling eventually becomes running.
Numerically it is a few hundred $MeV$ only, so it can not survive
in the UV (perturbative) limit, i.e., it can not be determined by
the theory in the UV region. It should come from the infrared
(IR) nonperturbative region. This clearly shows that this scale is
completely nonperturbative by origin and this conclusion is
strongly supported by the phenomenon of "dimensional
transmutation" [4].

Let us emphasize that the existence of the asymptotic scale parameter,
$\Lambda_{QCD}$, means that the full gluon propagator depends in general on this scale parameter, i.e., its behavior in the deep UV limit deviates from the free perturbative one (the above-mentioned running eefective coupling constant).
In other words, the perturbative dynamics even in massless QCD is nontrivial (scale violation, asymptotic freedom). Moreover, let us note in advance, that due
to the origin of the QCD scale parameter from the IR region, the asymptotics of
the full gluon propagator in the deep IR limit should deviate from the free one
as well. In other words, in the IR limit the QCD coupling is also running with
its, generally speaking, own scale parameter (see below).  In this connection,
let us note that if QCD itself is confining theory, then some characteristic scale (different, in general, from asymptotic scale parameter, $\Lambda_{QCD}$)
is very likely to exist.

 All this is in strict contradiction with quantum electrodynamics (QED), the theory of electromagnetic interactions. In QED the UV divergences are result of
the method of solution, the perturbation theory. Any exact solution to
QED, in principle, should not be asymptotically free. That is why neither the perturbative UV divergences nor the dimensional scale parameter necessarily should
appear in QED. So the UV divergences in QED are not physical, i.e., there is no
some nontrivial physics behind these divergences (unlike QCD, where asymptotic
freedom is behind the UV divergences). However, mathematically the UV renormalization program (in order to render the theory
finite) is, of course, the same as in QCD\footnote{Though historically the renormalization of QCD was done following the renormalization of QED.} but unlike
QCD, all the UV divergences in QED can be absorbed into the electron charge
and mass terms. So there is no running coupling constant and the photon propagator remains free one in the deep UV limit (no deviation from the perturbative free behavior in this limit). There is no doubt, that the drastic difference between these two theories is due to Abelian character of QED (no direct interaction between massless photons).

If the structure of QCD in the UV perturbative region is well
understood (because of asymptotic freedom, as it was briefly
discussed above), little is known about its structure in the
nonperturbative IR region. The main reason of this is, of course,
the phenomenon of confinement which remains the most important
unsolved yet problem in QCD. It is still remains to understand
whether confinement (in particular quark confinement) is some
additional external requirement in order to complete QCD to be a
self-consistent theory of strong interactions or QCD itself is
able to explain confinement and other nonperturbative phenomena
without involving some extra degrees of freedom. At the very
beginning of QCD, it was expressed an idea [5,6], that the
quantum excitations of the IR degrees of freedom because of
self-interaction of massless gluons only made it possible to
understand confinement as well as other nonperturbative effects,
for example such as dynamical chiral symmetry breakdown (DCSB).
In other words, the importance of the IR structure of the true
QCD vacuum was emphasized as well as its relevance to quark
confinement, DCSB and other way around.

In accordance with these ideas, let us suppose that QCD is intrinsically IR divergent theory as well.
 Again the physical reason of these divergences is self-interaction of massless gluons only.
  Asymptotic freedom is behind the UV perturbative divergences, while the nonperturbative
IR divergences themselves are behind confinement and other
nonperturbative effects in QCD. Thus in our picture they are more
fundamental than the perturbative ones. How to take them into
account? Of course, by the corresponding behavior of the full
gluon propagator in the deep IR domain since there is no hope for
an exact solution(s). In this connection, let us remind the
reader that the full dynamical information of any quantum gauge
field theory such as QCD is contained in the corresponding
quantum equations of motion, the so-called Schwinger-Dyson (SD)
equations for lower (propagators) and higher (vertices and
kernels) Green's functions. These equations should be also
complemented by the corresponding Slavnov-Taylor (ST) identities
which in general relate the above mentioned lower and higher
Green's functions to each other [1,7,8]. These identities are
consequences of the exact gauge invariance and therefore $"are \
exact \ constraints \ on \ any \ solution \ to \ QCD"$ [1].
Precisely this system of equations can serve as an adequate and
effective tool for the nonperturbative approach to QCD [1,9,10].
Among the above-mentioned Green's functions, the two-point
Green's function describing the full gluon propagator
\begin{equation}
iD_{\mu\nu}(q) = \left\{ T_{\mu\nu}(q)d(-q^2, \xi) + \xi L_{\mu\nu}(q) \right\} {1 \over q^2 },
\end{equation}
has a central place [1,7-11]. Here
$\xi$  is a gauge fixing parameter ($\xi = 0$,  Landau gauge) and
$T_{\mu\nu}(q) = g_{\mu\nu} - q_\mu q_\nu / q^2 = g_{\mu\nu } - L_{\mu\nu}(q)$.
Evidently, its free perturbative (tree level) counterpart is obtained by simply
setting the full gluon form factor $d(-q^2, \xi)=1$ in Eq. (1.1).
 In particular, the solutions to the above-mentioned SD equation for the full gluon propagator (1.1), are supposed to reflect the quantum structure of the QCD
ground state. It is a highly nonlinear integral equation containing
many different propagators, vertices and kernels [1,7-11]. For this reason     
it may have many (several) different exact solutions with different            
asymptotics in the deep IR
limit (the UV limit because of asymptotic freedom is uniquely determined),
describing thus many different types of quantum excitations of gluon field
configurations in the QCD vacuum. Evidently, not all of them can reflect the real structure of
the QCD vacuum. Let us emphsize that we even do not know the complete set of boundary conditions
in order to attempt to uniquely fix theory.

The deep IR asymptotics of the full gluon propagator can be
generally classified into the three different types: 1) the IR
enhanced (IRE) or IR singular (IRS) (for references see the next
section below), 2) the IR finite (IRF) [12] and 3) the IR
vanishing (IRV) ones [13]. As it was emphasized above, any
deviation in the behavior of the full gluon propagator in the IR
domain from the free one automatically assumes its dependence on
a scale parameter (at least one) responsible for the
nonperturbative dynamics, say, $\Lambda_{NP}$. This is very
similar to asymtotic freedom which requires  asymptotic scale
parameter, $\Lambda_{QCD}$ associated with nontrivial
perturbative dynamics (scale violation). Thus, any realistic
Ansatz (since exact solution(s) is(are) not known) for the full
gluon propagator in QCD will contain at least the two
above-mentioned scale parameters. In our approach, the
nonperturbative scale is more fundamental than the asymptotic
one, and if the above-mentioned Ansatz is realistic one, than, in
principle, it should give rise to the relation between them,
namely the asymptotic scale as $pure \ number$ times the
nonperturbative scale. The main purpose of this article is
precisely to establish such relation within our approach to
nonperturbative QCD and its ground state.

However, a few general remarks are in order. The nonperturbative
QCD vacuum is a very complicated medium and its dynamical and
topological complexity [1,14-17] means that its structure can be
organized at various levels (classical and quantum) and it can
contain many different components and ingredients which
contribute to the truly nonperturbative vacuum energy density ,
one of the main characteristics of the QCD ground state. The
classical component of the vacuum energy density is determined by
the topologically nontrivial, instanton-type fluctuations of
gluon fields configurations in the QCD vacuum which are
nonperturbative weak-coupling limit solutions to the classical
equation of motion in Euclidean space. Numerically this
contribution is estimated by the random and interacting instanton
liquid models (RILM and IILM) of the QCD vacuum (see Ref. [18]
and references therein).

Here we are going to discuss the quantum part of the vacuum energy density which
is determined by the effective potential approach for composite operators
[19,20]. It allows us to investigate the nonperturbative QCD vacuum, since in the
absence of external sources the effective potential is nothing
but the vacuum energy density. It gives the vacuum energy density in the form of the
loop expansion where the number of the vacuum loops (consisting of the confining quarks
and nonperturbative gluons properly regularized with the help of
ghosts) is equal to the power of the Plank constant, $\hbar$.

 It is well known, however, that the vacuum energy density is generally badly divergent
  in quantum field theory, in particular QCD (see, for example, the discussion
  given by Shifman in Ref. [14]). The main problem
thus was how to extract the truly nonperturbative vacuum energy
density which should be finite, automatically negative without
imaginary part (stable vacuum). This problem was addressed and
resolved in our recent publications in Ref. [21] for covariant
gauge QCD and in Ref. [22] for axial gauge QCD. Though the truly
nonperturbative vacuum energy density is important in its own
right (it is the bag constant, apart from the sign, by definition
and precisely it is related to the gluon and quark condensates,
topological susceptibility, etc. (see Refs. [23-26] and references
therein)), here we are going to investigate the general structure
of the Yang-Mills (YM) vacuum within our approach.

In the next section II we will formulate our main dynamical
assumption which lies in the heart of our approach to
nonperturbative QCD in general and to its ground state in
particular.  Then in section III we will apply to the
above-mentioned approach for composite operators in order to
analyze the vacuum structure. In section IV we present our main
result while section V is devoted to discussion. Our conclusions
are given in section VI.

\section{The IRE gluon propagator. ZME quantum model}

Today there are no doubts left that the dynamical mechanisms
of the important non-perturbative quantum phenomena such as quark confinement and
dynamical (or equivalently spontaneous) chiral symmetry breaking (DCSB) are closely
related to the complicated topologically nontrivial structure of
the QCD vacuum [1,14-17]. On the other hand, it also becomes clear that the nonperturbative
IR dynamical singularities, closely related to the nontrivial vacuum
structure, play an important role in the large distance behaviour
of QCD [5,6]. For this reason, any correct nonperturbative model of quark confinement
and DCSB necessarily turns out to be a model of the true QCD vacuum
and the other way around.

Our model of the true QCD ground state is based on the existence and importance
of such kind of the nonperturbative, quantum excitations
of the gluon field configurations (due to self-interaction of massless gluons only,
i.e., without explicit involving some extra degrees of freedom) which can
be effectively correctly described
by the $q^{-4}$ behavior of the full gluon propagator in the deep IR domain (at
small $q^2$) [27,28].
Thus our main Ansatz for the full gluon form factor in Eq. (1.1) in the whole
momentum range $[0, \infty)$ is

\begin{equation}
d(-q^2, \xi) = {\mu^2 \over (-q^2)} + {c  \over \ln(- q^2 / \Lambda^2_{YM})}.
\end{equation}
Evidently, the UV piece is nothing but the running coupling
constant $\bar g^2(-q^2)$, which is the leading order solution to
the renormalization group equations [1] and $c=( 16 \pi^2 /11)$
is the inverse of the first coefficient of the $\beta$-function
in the case of YM gluon fields (pure gluodynamics). In the
above-mentioned Ref. [1] it is denoted as $2/b_0$. The flavorless
QCD asymptotic scale parameter is denoted as $\Lambda_{YM}$. In
the weak coupling limit ($ - q^2 \rightarrow \infty$) precisely
the second term becomes dominant (asymptotic freedom).

On the other hand, we obtain the generally accepted form of the IR singular
asymptotics for the full gluon propagator (for some references see below)
\begin{equation}
D_{\mu\nu}(q) \sim (q^2)^{-2} , \qquad q^2 \rightarrow 0,
\end{equation}
which may be equivalently referred to as the strong coupling regime [1].
It describes zero momentum modes enhancement (ZMME) dynamical effect in QCD
at large distances. In our previous publications and in future we prefer to use
simply ZME (zero modes enhancement) since we work
in momentum space always. This is our primary dynamical assumption in this section. The main problem
with this strong singularity is its correct treatment by the dimensional
regularization method [29] within the distribution theory [30], which precisely
was one of highlights of our previous publications [27,28] (see also
Refs. [31,32]).
There exist many arguments in favor of this behaviour. Let us remind the
reader a few of them which are the most important in our opinion.

a). Precisely this strongly singular behaviour of the full gluon propagator in
the IR domain leads to the area law for static quarks (indicative of confinement)
within the Wilson loop approach [33].

b). The cluster property of the Wightman functions in QCD fails and this allows
such singular behavior like (2.2) for the full gluon propagator in the deep IR
domain [34].

c). After the pioneering papers of Mandelstam
in the covariant (Landau) gauge [35] and Baker, Ball and Zachariasen in the axial gauge [36],
the consistency of the singular asymptotics (2.2) with direct solution to the SD equation
for the full gluon propagator in the IR domain was
repeatedly confirmed (see for example Refs. [9,10, 37,38] and references therein).

d). Moreover, let us underline that
without this component in the decomposition of the full gluon propagator in continuum theory
it is impossible to "see" linearly rising potential between heavy
quarks by lattice QCD simulations [39] not involving
some extra (besides gluons and quarks) degrees of freedom.
This should be considered as a strong lattice evidence (though not direct) of the existence
and importance of $q^{-4}$-type excitations of gluon field configurations in the QCD vacuum.
There also exists direct lattice evidence that the zero modes are enhanced in the full gluon
propagator indeed [40].

e). Within the distribution theory [30] the structure of the nonperturbative IR
singularities in four-dimensional Euclidean QCD is the same as in two-dimensional QCD,
 which confines quarks at least in the large $N_c$ limit [41]. In this connection, let us
 note that $q^{-4}$ IR singularity is the simplest nonperturbative power singularity
 in four-dimensional QCD as well as $q^{-2}$ IR singularity is the simplest
 nonperturbative power singularity in two-dimensional QCD. The QCD vacuum is
 much more complicated medium than its two-dimensional model, nevertheless,
 the above-mentioned analogy is promising even in the case of the nonperturbative
 dynamics of light quarks.

f). Some classical models of the QCD vacuum also invoke $q^{-4}$ behavior of the gluon fields
in the IR domain. For example, it appears in the QCD vacuum as a
condensation
of the color-magnetic monopoles (QCD vacuum is a chromomagnetic superconductor)
proposed by Nambu, Mandelstam and 't Hooft and developed by Nair and Rosenzweig
(see Ref. [42] and references therein. For recent developments in this model
 see Di Giacomo's contribution in Ref. [14]) as well as in the classical mechanism
of the confining medium [43] and in effective theory for the QCD vacuum proposed in Ref. [44].

g). It is also required to derive the heavy quark potential within
the recently proposed exact renormalization group flow equations approach [45].

h). It has been shown in our papers that the singular behavior
(2.2) is directly related to light quarks confinement and DCSB
[27,28].

i). These excitations are also topologically nontrivial since they lead to the nontrivial
 (truly nonperturbative) vacuum energy density which is finite, automatically negative and it
  has no imaginary part (stable vacuum) [21,27,28].

j). Moreover, they nicely saturate the phenomenological values of the topological susceptibility,
the mass of the $\eta'$ meson in the chiral limit and the gluon condensate [25,26].

Thus we consider our main Ansatz (2.1) as physically
well-motivated. The first term in the decomposition (2.1) is the
truly nonperturbative part since it vanishes in the perturbative
limit ($\mu^2 \longrightarrow 0$), when the perturbative phase
survives only and simultaneously it correctly reproduces the deep
IR asymptotics of the full gluon propagator.

\section{The effective potential for composite operators }

 In this section we are going to analytically investigate the quantum structure
of the YM vacuum which is due to $q^{-4}$-type nonperturbative, topologically nontrivial
 excitations of the gluon field configurations there complemented by the renormalization
 group improvements for the UV region, (2.1).
Let us start from the gluon part of the vacuum energy density
which to leading order (log-loop level $\sim \hbar$)\footnote{Next-to-leading and higher terms
(two and more vacuum loops) are suppressed
by one order of magnitude in powers of $\hbar$ at least and are left for consideration elsewhere.}
is given by the effective potential for composite operators [19] as follows

\begin{equation}
V(D) =  { i \over 2} \int {d^nq \over {(2\pi)^n}}
 Tr\{ \ln (D_0^{-1}D) - (D_0^{-1}D) + 1 \},
\end{equation}
where $D(q)$ is the full gluon propagator (1.1) and $D_0(q)$ is its
free perturbative (tree level) counterpart. Here and below the traces over
space-time and color group indices are understood.
The effective potential is normalized as $V(D_0) = 0$, i.e., free perturbative
vacuum is normalized to zero. In order to evaluate the effective potential (3.1)
 we use the well-known expression,

\begin{equation}
 Tr \ln (D_0^{-1}D) = 8 \times \ln det (D_0^{-1}D) =
 8 \times 4 \ln \left[ {3 \over 4 }d(-q^2) + {1 \over 4 } \right].
\end{equation}
It becomes zero (in accordance with the above mentioned normalization condition)
 when the full gluon form factor is replaced by its free counterpart. The composition (3.2)
  does not explicitly depend on a gauge choice.

 Going over to four ($n=4$) dimensional Euclidean space in (3.1-3.2),
 on account of (2.1), by introducing the appropriate dimensionless momentum
 variable as $z= q^2 / \mu^2$ and after doing some algebra, we finally obtain
 the following expression for the vacuum energy density, $\epsilon_g = V(D)$,

\begin{equation}
\epsilon_g = {1 \over \pi^2} q_0^4 z_0^{-2} I_g(z_0; b),
\end{equation}
where

\begin{equation}
I_g (z_0; b) = \int \limits_0^{z_0} dz\, z\, \Bigl\{ \ln z + {3 \over 4z} -a +
 {3c \over 4 \ln bz} - \ln [3+z+ {3cz \over \ln bz }] \Bigr\} = I^r_g (z_0; b) + I^s_g (z_0; b),
\end{equation}
and

\begin{eqnarray}
I^r_g (z_0; b) &=&  {1 \over 2} z_0^2 \ln \bar b z_0
- {1 \over 4} z^2_0 + {3 \over 4} z_0 - I(z_0; b) \nonumber\\
I^s_g (z_0; b) &=& -  {\bar a \over 2} z^2_0 + {3c \over 4b^2}li(bz_0)^2 +
{1 \over 2} z_0^2 \ln \ln b z_0 + {1 \over 2 b^2} Ei(1, - 2 \ln bz_0),
\end{eqnarray}
with $\bar a = a + \ln \bar b$ and $a= (3/4) - 2 \ln 2$ while
$\bar b = \ln bz_0 + 3 c$. In these integrals $z_0 = q_0^2 /
\mu^2$ while $b= \mu^2 / \Lambda^2_{YM}$ and $li(x)$ is the
logarithm-integral function. Let us note in advance, that here
and everywhere below the combination $bz_0 = q_0^2 /
\Lambda^2_{YM} = x_0$ is fixed (constant) when $q_0$ itself is
fixed, i.e., before going to infinity at final stage. Thus, the
parameter $x_0$ is the truly dimensionless UV cutoff since when
the UV cutoff in momentum space $q_0^2$ goes to infinity ($q_0^2
\rightarrow \infty$) then it certainly goes to infinity as well.
At the same time, the variable $z_0$ only looks like as the
dimensionless UV cutoff. In the limit $q_0^2 \rightarrow \infty$
it might be finite since the nonperturbative scale, $\mu^2$, can
be affected by the UV divergences as well ("bare" parameter).
Precisely this takes place within our approach in this case (see
below). In fact, variable $z_0$ determines the perturbative limit
($z_0 \rightarrow \infty$) at fixed $q_0^2$, when the
perturbative phase survives only in our model ($\mu \rightarrow
0$). $Ei(1, - 2 \ln bz_0)$ is the exponential-integral function
with unphysical singularity at $bz_0=1$ as well as the preceding
term has. In the compositions shown in (3.5), the corresponding
integral is defined as follows

\begin{equation}
I(z_0; b) = \int \limits_0^{z_0} dz\, z\, \ln [(3+z) \ln bz + 3cz].
\end{equation}

 The effective potential at the log-loop level for the ghost degrees of freedom
is

\begin{equation}
V(G) = - i \int {d^np \over {(2\pi)^n}}
Tr\{ \ln (G_0^{-1} G) - (G_0^{-1} G) + 1 \},
\end{equation}
where $ G(p)$ is the full ghost propagator and $G_0(p)$ is its
free perturbative counterpart. The effective potential $V(G)$ is normalized
as $V(G_0) = 0$. Evaluating the ghost term $\epsilon_{gh} = V(G)$ in (3.7), we
obtain $\epsilon_{gh} = \pi^{-2} k_0^4 y_0^{-2}I_{gh}(y_0; b)$,
where the integral $I_{gh}(y_0; b)$ depends on the ghost propagator,
which remains arbitrary (unknown) within our approach and $y_0 = k_0^2 / \Lambda_{NP}^2$.

 In principle, we must sum up all contributions in order to obtain total vacuum
energy density (the confining quark part of the vacuum energy density is not considered here). However, approximating the full gluon
propagator by the sum of its deep IR (confinement) and UV asymptotics in (2.1),
the nonzero terms appear in (3.4) which precisely violates the above mentioned
normalization condition of the free perturbative vacuum to be zero when the parameter $b$ goes to zero ($\mu \rightarrow 0, z_0 \rightarrow \infty$).
In this case the perturbative phase only remains within our model. These terms
are
collected in the expression (3.5) for $I^s_g (z_0; b)$. It also contains two terms which suffer from unphysical singularity
- Landau pole - at $x_0=1$. This should be traced back to (2.1) when $ - q^2 = \Lambda^2_{YM}$ since we use it in the whole range. These terms should be subtracted from the gluon part of the vacuum energy density. Using the arbitrariness
of the ghost term (3.7), this can be done by imposing  the following condition,
$\Delta = I^s_g (z_0; b) +  I_{gh}(y_0; b) = 0$ (recalling that $\Lambda^4_{YM} = k_0^4 y_0^{-2}= q_0^4 z_0^{-2}$). We regularize the gluon contribution to the vacuum energy density by subtracting unwanted terms by
means of the ghost contribution, i.e., it plays the role of the corresponding counter term. So we define, $ \epsilon_g + \epsilon_{gh} = \epsilon^{reg}_g = \pi^{-2} q_0^4 z_0^{-2}I^r_g (z_0; b)$, on account of $\Delta=0$. Thus our regularization procedure
is in agreement with general physical interpretation of ghosts to cancel the effects of unphysical degrees of freedom of the gauge bosons [1,46].

 Let us introduce the effective potential $\bar \Omega_g$ at fixed $q_0$ as follows [21,28]

\begin{equation}
\bar \Omega_g \equiv \bar \Omega_g (z_0; b) = { 1 \over  q_0^4} \epsilon_g^{reg} = {1 \over 2 \pi^2} \times z_0^{-2}
\Bigl\{ z_0^2 \ln \bar b z_0
- {1 \over 2} z^2_0 + {3 \over 2} z_0 - 2 I(z_0; b) \Bigr\},
\end{equation}
where the integral $I(z_0; b)$ is given in (3.6). First of all, it is always useful to factorize the scale dependence in the regularized and subtracted vacuum
energy density by means of the above-introduced effective potential at fixed scale.
On the other hand, it makes it possible to investigate the structure of the vacuum within our approach since this quantity is finite when the UV cutoff goes to infinity (see below) as well as it is free from unphysical singularities (see
above). These two remarks emphasize the importance of the effective potential (3.8). The asymptotic behavior of the effective potential (3.8) depends
on the asymptotic properties of the integral $I(z_0; b)$, (3.6). It is almost
obvious that its asymptotics at $z_0 \rightarrow 0, \infty$ (and consequently at $b \rightarrow 0, \infty$) to-leading order can be easily evaluated analytically by replacement $\ln bz$ by $\ln bz_0$ since logarithm is a very slowly varing function. Omitting all intermediate analytical calculations, one obtains

\begin{equation}
I(z_0; b) = {1 \over 2} z_0^2 \ln (\bar b z_0) - {1 \over 4} z^2_0 + {3 \over 2 \bar c} z_0 - {9 \over 2 \bar c^2} \ln \Bigl\{ 1 + {\bar c z_0 \over 3}\Bigr\} + {1 \over 2} z_0^2 \ln \Bigl\{1 + {3 \over \bar c z_0 } \Bigr\},
\end{equation}
where $\bar c = 1 + (3c / \ln bz_0)$. Thus in the perturbative limit $z_0 \rightarrow \infty$ one has

\begin{equation}
\bar \Omega_g(z_0; b) \sim_{z_0 \rightarrow \infty} - {9 \over 4 \pi^2} {\ln bz_0 - c \over \ln bz_0 + 3c }z_0^{-1},
\end{equation}
which clearly shows that in order for the effective potential (3.8) as a function of $z_0$ to approach zero from below in the perturbative limit, one needs to
set

\begin{equation}
\ln x_0 = \lambda c, \qquad \lambda > 1,
\end{equation}
recalling that $bz_0 = x_0$ by definition. The expression (3.11) is nothing but
useful parametrization of the UV divergences in terms of the parameter $\lambda$, which will go to infinity at the final stage. At the same time, at zero ($z_0 \rightarrow 0$) the effective potential (3.8) behaves as $\sim z_0^{-1}$. Thus as a function of $z_0$ it
has a minimum if the condition (3.11) takes place which is the case indeed
($\lambda$ goes to infinity at final stage).

The minimization of the effective potential (3.8) with respect to $z_0$ (at fixed $bz_0$), $ \partial \bar \Omega_g (z_0; b) / \partial z_0 = 0$, yields

\begin{equation}
z_0^2 + 4 I_2(z_0; b) = {3 \over 2} z_0 + 2 z_0^2 \ln [(3 + z_0) \ln bz_0 + 3cz_0].
\end{equation}
It is easy to show that to-leading order in the  $\lambda \rightarrow \infty$ limit, it can be analytically evaluated as follows:

\begin{equation}
(\lambda - 1)(\lambda + 3) z_0 = 4 \lambda^2  \ln \Bigl\{ 1 + {(\lambda +3) z_0 \over 3 \lambda} \Bigr\}.
\end{equation}
In the derivation of this expression we have already used expression for $\bar c$ (see text after Eq. (3.9)) and (3.11). Its solution in the $\lambda \rightarrow \infty$ limit is

\begin{equation}
z_0^{min} (\lambda \rightarrow \infty) = \Bigl\{ { x_0 \over b} \Bigr\}_{min} ( \lambda \rightarrow \infty) = 2.2,
\end{equation}
i. e. the point $z_0^{min}= 2.2$ is the point of accumulation of
minimums in the UV limit of the effective potential (3.8). This
effect shows that the parameter $b$ goes to infinity in the same
way as the UV cut-off $x_0$ in order for their ratio to go to the
above displayed finite limit. This means that initial mass scale
parameter $\mu$ is a "bar" one. It depends on the UV cut-off
$\lambda$, i.e., $\mu = \mu (\lambda)$ and consequently $b =
b(\lambda)$\footnote{Let us note that it does not make any sense
to introduce finite (from the very beginning) the nonperturbative
mass scale parameter. The problem is that in any case it will be
affected by the UV divergences since we have explicitly kept the
UV term in Eq. (2.1) (did not subtract it). At the same time, the
asymptotic scale parameter, $\Lambda_{YM}$, is finite (from the
very beginning) by definition.}. In what follows we will
introduce the UV renormalized $\mu$ and consequently the UV
renormalized $b$. To-leading order in the $\lambda \rightarrow
\infty$ limit the effective potential (3.8), on account of the
"stationary" condition (3.13), becomes

\begin{equation}
\bar \Omega_g(\lambda, z_0) = { 1 \over 2 \pi^2} \Bigl\{ {3 \over 4}z_0^{-1}-
\ln (1 + {3 \lambda \over z_0 (\lambda +3)}) \Bigr\}.
\end{equation}

\section{The YM vacuum structure}

Evaluating the effective potential $\Omega_g(\lambda, z_0^{min}(\lambda))$ (3.15) numerically in the vicinity of the point of accumulation of the minimums, $z_0^{min}(\lambda \rightarrow \infty) \rightarrow 2.2$ (see (3.14)), one obtains
$\Omega_g(\lambda \rightarrow \infty) = - 0.0263$, i.e., it is finite in this limit. Recalling that $\epsilon_g^{reg} = \epsilon_g^{reg}(\lambda)$, then from
Eq.(3.8) it follows

\begin{equation}
\epsilon_g^{reg}(\lambda) = - 0.0263 q_0^4 (\lambda).
\end{equation}
On the other hand,

\begin{equation}
q_0^4(\lambda) = x_0^2(\lambda) \Lambda^4_{YM}= e^{2 \lambda c} \Lambda^4_{YM},
\end{equation}
by definition and because of (3.11). Thus the vacuum energy becomes

\begin{equation}
\epsilon_g^{reg}(\lambda) = - 0.0263 \Lambda^4_{YM} \times e^{2 \lambda c}
\end{equation}
and it is badly divergent in the UV limit ($\lambda \rightarrow \infty$) because of the exponential $\lambda$-factor. This result should be expected since we
have explicitly kept (did not subtract) the UV peace in Eq. (2.1) (for general
discussion see Ref. [21]).

However, despite badly divergent vacuum energy density, our method makes it possible to establish a finite relation between two scale parameters of our model.
Indeed, as was mentioned above, the parameter $b$ diverges exactly as the UV cutoff $x_0 = e^{\lambda c}$ (see Eq. (3.11)) itself in order to get final result
in the $\lambda \rightarrow \infty$ limit. So let us introduce now the UV
finite $b$ and $\mu$ as

\begin{equation}
\tilde{b} = b(\lambda) e^{- \lambda c}
\end{equation}
and

\begin{equation}
\Lambda_{NP} =\mu (\lambda) e^{- \lambda c/ 2 },
\end{equation}
respectively. Then from Eq. (3.14) one obtains

\begin{equation}
z_0^{min} = { x_0 \over b (\lambda)} = { e^{ \lambda c } \over  \tilde{b}
e^{\lambda c}} = \tilde{b}^{-1} = 2.2
\end{equation}
in the $\lambda \rightarrow \infty$ limit. On the other hand, by definition

\begin{equation}
\Lambda^2_{YM} = {\mu^2 (\lambda) \over b(\lambda)} = {\Lambda_{NP}^2  e^{ \lambda c } \over \tilde{b} e^{ \lambda c }} = {\Lambda_{NP}^2 \over \tilde{b}}
\end{equation}
in the same limit. This finally leads to

\begin{equation}
\Lambda_{YM} = 1.48324 \times \Lambda_{NP}.
\end{equation}

A few remarks are in order. Let us underline that the above-displayed numerical
result (4.8) was obtained in a gauge invariant way, i.e., this number did not explicitly depend on a gauge choice. Also this result will not be
changed in the next-to-leading order approximation to the running coupling constant in the UV limit since they were obtained to-leading dominant order limit.
Evidently, it should not, of course, depend on how one introduces the UV cutoff by making use of different scale parameters. For example, it can be
defined as $x_0 = q_0^2 / \Lambda_{YM}^2$ (from the very beginning) while the parameter $b$ remains the same. Then the minimization of the effective potential
$\Omega_g(x_0; b) = \epsilon_g^{reg} (x_0; b) \Lambda^{-4}_{YM}$ with respect to $b$ finally yields the
same numerical relation between the nonperturbative and asymptotic scales, (4.8) as it should be in principle (for details see Ref. [47]). It is worth noting
that the minimization with respect to $x_0$ itself in this case leads to trivial zero (see Ref. [21]).

\section{Discussion}

The main problem now is the numerical value of the YM asymptotic scale parameter, $\Lambda_{YM} \equiv \Lambda$ (for brevity and further aims), since it can
not be experimentally determined in the YM theory alone. So its numerical value
is to be taken from lattice approach or phenomenology. Unfortunately, lattice estimates of the QCD flavorless asymtotic scale parameter strongly depend on the
renormalization
scheme chosen for calculation. In Ref. [48] in the $\overline{MS}$ scheme it is
expressed in terms of Sommer's scale $r_0$, namely $\Lambda_{\overline{MS}} = 0.636(54)/ r_0$. Than Eq. (4.8) becomes
$\Lambda_{NP} = 0.4288 r_0^{-4}$. In physical units, using  $r_0 = 0.5
\ fm$, one numerically gets, $\Lambda_{NP} = 171.5 \ MeV$. However, as
it was emphasized in the above mentioned Ref. [48], the conversion to physical
units is rather ambiguous in the pure gauge theory. Approximately the same numerical result for $\Lambda_{\overline{MS}}$ in terms of the string tension was previously obtained in Ref. [49]. In Ref. [50] the YM scale parameter was calculated first in the $\tilde{MOM}$
scheme and then it was matched to the $\overline{MS}$ scheme. The corresponding
numerical results (here and everywhere below up to central values, for simplicity) are: $\Lambda_{\tilde{MOM}} = 0.88 \ GeV$ and $\Lambda_{\overline{MS}} = 0.31 \ GeV$. Substituting these values into the Eq. (4.8), one arrives at the following estimates, $\Lambda_{NP}^{\tilde{MOM}} = 593.3 \ MeV$ and $\Lambda_{NP}^{\overline{MS}} = 210 \ MeV$.
The most recent lattice calculation of the flavorless QCD asymptotic scale parameter is being presented in Ref. [51] are:
$\Lambda_{MOM} = 0.361(6) \ GeV$ and
$\Lambda_{\tilde{MOM}} = 0.349(6) \ GeV$ while $\Lambda_{\overline{MS}} = 0.412 \ GeV$. Substituting this result into the Eq. (4.8) for example for the last value, one gets, $\Lambda_{NP}^{\overline{MS}} = 277.7 \ MeV$.

 In Ref. [52] the vacuum energy density of the covariant model of Euclidean QCD
with $N_f=3$ quark flavors on a finite $D$-dimensional symmetric hypertorus was
calculated. Adjusting the recently calculated value of the gluon condensate by
Narison [53] and the numerical value of $\alpha_s$ from $\tau$ decay, in the above-mentioned paper [50] it was estimated that $\Lambda^{(3)}_{\overline{MS}} \sim 500 - 600 \ MeV$ (compatible with the standard value of the gluon condensate [54]).
From the renormalization group equations, however, it is well-known that the numerical value of the
flavorless QCD asymptotic scale parameter should be bigger
than its full QCD counterparts, for example, the above estimated $\Lambda^{(3)}_{\overline{MS}}$. Thus roughly estimating  $\Lambda_{YM} \sim 600 -700 \ MeV$
in Eq. (4.8), one obtains $\Lambda_{NP} \sim 404.5 - 472 \ MeV$. Contrary to the above-discussed cases, in Refs. [27,28] the nonperturbative scale parameter
was calculated itself in chiral QCD within our approach. Because of the scale-setting scheme in order to relate it to a good physical observeble from full QCD
, its numerical value depends on a bounds for the
pion decay constant in the chiral limit. The corresponding bounds for the nonperturbative scale parameter, we have obtained, were: $540.6 \leq \bar \mu \leq 578.4 \ MeV$. Then from Eq. (4.8) it immediately follows
 $801.8 \leq \Lambda_{YM} \leq 858 \ MeV$ by identifying $\bar \mu$ with $\Lambda_{NP}$ for simplicity. Here we are observing the standard situation when estimates from phenomenology (continuum theory) are usually bigger
than lattice estimates of the same scale quantities.

In any case, our solid result is the relation (4.8) which is to be better understood as the ratio between two scale parameters from lattice simulation point of view. In this way, in principle, it should be free from lattice artifacts since it does not require the conversion to physical unuts which still remains rather ambiguous in the pure gauge theory as it was mentioned above.

\section{Conclusions}

In summary, we have formulated new quantum model of the QCD vacuum using the   
effective potential approach for composite operators. It is based on the       
existence and importance
of such kind of the nonperturbative, topologically nontrivial excitations of
gluon field configurations (due to self-interaction of massless gluons only,   
i.e., without explicit involving some extra degrees of freedom) which can be
effectively correctly described by the $q^{-4}$-type behaviour of
the full gluon propagator in the deep IR domain. The UV part of
the full gluon propagator was approximated by asymptotic freedom
to-leading order log term of the running coupling constant. By
explicit retaining the UV tail in the decomposition of the full
gluon propagator (2.1), the vacuum energy density remains badly
divergent, of course (for general discussion see again Refs.
[21,22] and Shuryaks's review in Ref. [55]). However, despite
this, we have formulated a method to establish a finite (in the
UV limit) relation (4.8) between two scale parameters of our
model. We propose to minimize the properly regularized and
subtracted effective potential at a fixed scale (3.8) as a
function of the variable $z_0$. It has a clear physical meaning
when it goes to infinity then the perturbative phase survives
only within our approach to nonperturbative QCD. This is possible
to do because the above-mentioned effective potential is finite
in the UV limit within our model, i.e., this is a specific property of our     
Ansatz for the full gluon propagator (2.1).                                    
Equivalently, one can minimize
the auxiliar (again properly regularized and subtracted)
effective potential as a function of the parameter which has a
clear physical meaning. When it is zero then again the
perturbative phase survives only.                                              
  
For the first time, we have
expressed the asymptotic scale parameter as $pure \ number$ times
the nonperturbative scale, namely (4.8). We don't know any other
YM quantum/classical vacuum model which possesses the same
property. This can be considered as one more argument in favor of
our model. In principle, in QCD one of the final goals is
precisely to express all scale parameters (for example, hadron
masses) as $pure \ numbers$ times some characteristic scale of
the theory. Despite the perturbative vacuum cannot be the QCD true        
ground state [56], nevertheless the                                            
existence of such kind of relation is a
manifestation that "the problems encountered in perturbation
theory are not mere mathematical artifacts but rather signify
deep properties of the full theory" [57]. In support of this statement, let us 
underline one more that the running coupling with scaling violations           
not only in the UV but in the IR as well is the most important feature         
of QCD which dratically differs this theory from QED. The former is certainly  
related to asymptotic freedom while the latter should be apparently related to
confinement, in particular quark confinement. Behind both types of scaling violation (in general, with their own scale parameters) is the self-interaction of 
massless gluons only within our approach to nonperturbative QCD in general and 
to its ground state in particular. Thus it suggests a unified description of both phenomena, so such kind of relation between two scale parameters, in principle, should be expected. At the same time, it does not mean that they are       
dependent from each other. It means rather that there exists the procedure     
(method) how to establish connection between them. 
Precisely this point we wanted to emphasize in the Introduction.

\acknowledgements

The author would like to
thank M. Polikarpov, M. Chernodub, A. Ivanov and especially V.I. Zakharov for useful and critical discussions as well as A.T. Filippov for corespondence. It is a pleasure to thank Gy. Kluge for many remarks and help.This
work was supported by HAS-JINR collaboration agreement.


\vfill

\eject

\end{document}